\documentclass[sigconf,nonacm]{acmart}

\usepackage{booktabs}
\usepackage{tablefootnote}

\AtBeginDocument{%
  \providecommand\BibTeX{{%
    \normalfont B\kern-0.5em{\scshape i\kern-0.25em b}\kern-0.8em\TeX}}}

\setcopyright{acmlicensed}
\copyrightyear{2024}
\acmYear{2024}
\acmDOI{XXXXXXX.XXXXXXX}

\acmConference[Conference acronym 'XX]{Make sure to enter the correct
  conference title from your rights confirmation email}{June 03--05,
  2018}{Woodstock, NY}
\acmISBN{978-1-4503-XXXX-X/18/06}

\begin{document}

\title{Assessing the Performance of Human-Capable LLMs - Are LLMs Coming for Your Job?}

\author{John Mavi}
\affiliation{%
  \institution{Luxembourg Tech School A.s.b.l.}
  \country{Luxembourg}
}
\email{johnmavi08@gmail.com}

\author{Nathan Summers}
\affiliation{%
  \institution{Luxembourg Tech School A.s.b.l.}
  \country{Luxembourg}}
\email{nathan.summers@men.lu}

\author{Sergio Coronado}
\affiliation{%
  \institution{Luxembourg Tech School A.s.b.l.}
  \country{Luxembourg}}
\email{sergio.coronadoarrechedera@men.lu}


\begin{abstract}
    The current paper presents the development and validation of \textbf{SelfScore}, a novel benchmark designed to assess the performance of automated Large Language Model (LLM) agents on help desk and professional consultation tasks. Given the increasing integration of AI in industries, particularly within customer service, SelfScore fills a crucial gap by enabling the comparison of automated agents and human workers. The benchmark evaluates agents on problem complexity and response helpfulness, ensuring transparency and simplicity in its scoring system. The study also develops automated LLM agents to assess SelfScore and explores the benefits of Retrieval-Augmented Generation (RAG) for domain-specific tasks, demonstrating that automated LLM agents incorporating RAG outperform those without. All automated LLM agents were observed to perform better than the human control group. Given these results, the study raises concerns about the potential displacement of human workers, especially in areas where AI technologies excel. Ultimately, SelfScore provides a foundational tool for understanding the impact of AI in help desk environments while advocating for ethical considerations in the ongoing transition towards automation.
\end{abstract}


\maketitle

\section{Introduction}
Automation and efficiency have long been inextricable pursuits of scientific research and societal advancement. During the Industrial Revolution of the 18th and 19th centuries, the development and proliferation of steam power and advancements in production techniques facilitated the widespread establishment of the emerging manufacturing economy, as mass production became easier and more efficient. Later, starting in the 1940s and 1950s, the transition from analog to digital technologies began, gaining momentum in the following decades. With it, unprecedented computational power became more widely available, trivializing previously labor-intensive (or even resource-prohibitive) tasks.

Now, the precipice of a new technological revolution looms ahead. With the increasing availability of massive amounts of data, Artificial Intelligence (AI) technologies are experiencing an explosion in their efficacy, their applicability, and their accessibility. ChatGPT was among the fastest growing products in history~\cite{hu_chatgpt_2023}, and retains a large daily active userbase today, two years later~\cite{noauthor_chatgptcom_nodate,noauthor_chatopenaicom_nodate}. AI is being increasingly integrated into social media~\cite{noauthor_meta_2023,morera_metas_2024,noauthor_musk_2023}, healthcare~\cite{perrone_will_2024,nellis_mayo_2024}, business and enterprise~\cite{noauthor_what_nodate,noauthor_what_2023,noauthor_ai_nodate}, and transportation~\cite{breunig_building_nodate,gonzalez_council_nodate,early_can_2023}, and these trends show no signs of slowing. Although AI is still in its infancy, efficiency gains are already being observed~\cite{noy_experimental_2023,gao_ai-driven_2023}.

Technology-driven efficiencies have historically had significant impacts on contemporary society, particularly within the labor market. Mechanization, initiated during the Industrial Revolution, reduced the labor demand in agriculture, leading to a long-term decrease in the number of farmers~\cite{cartwright_agriculture_nodate}. The advent of digital technologies led to a transition within the workforce, emphasizing the prioritization of knowledge-based roles~\cite{buchanan_digital_nodate,mandl_employment_nodate}. With AI, the potential ramifications stand to be much more substantial. The question then naturally arises, how will these revolutionary technologies shape the future of labor, as those preceding it did?

Automated agents are of particular import to this question. These are software systems that autonomously perform tasks or make decisions based on predefined rules or learning algorithms. They operate in a wide range of domains, from customer service or financial trading to programming and development. They are designed to simulate human decision-making without direct intervention. Some agents can incorporate the capability to autonomously carry out actions based on their simulated decision-making~\cite{pogla_auto-gpt_2024}.

Because automated agents do not rely on direct intervention, coupled with the fact that machines are capable of processing and outputting data more efficiently than humans, these agents stand to drastically change the nature of work. Automated agents might be deployed to oversee rote tasks, such as reporting and administrative work that currently requires human oversight, thereby permitting humans to instead work on more complex, creative, or cognitively demanding tasks for which AI is not currently suited. By streamlining certain processes, resources can be redistributed to significantly increase efficiency throughout the working world.

Key to current implementations of automated agents are Large Language Models (LLMs), as they permit ordinary people to meaningfully interact with computers using tools they already possess: language. LLMs are advanced AI systems trained on vast amounts of text data to understand and generate human-like language. They use deep learning techniques, such as neural networks, to predict and produce coherent text based on a given input, making them useful for a wide array of tasks like translation, summarization, and conversation~\cite{noauthor_what_2023-1,noauthor_introduction_nodate}. Consequently, they often form the foundation upon which automated agents, or automated LLM agents, are built.

This permits individuals to describe a task to an automated agent using natural language. The agent is then able to leverage its constituent subsystems to autonomously carry out this task and provide its user with an output that is not restricted to natural language text. This flexibility, multi-modality, and versatility distinguishes automated LLM agents from their foundational LLMs.

However, many LLMs are trained on broad, non-domain-specific text in order to maximize the quantity of training data and provide the LLM with the most comprehensive understanding of how natural language works. Retrieval-Augmented Generation (RAG) can be used to mitigate this shortcoming. RAG is an advanced approach in natural language processing (NLP) that combines retrieval of relevant information from external sources with generation of new text based on that information~\cite{lewis_retrieval-augmented_2021,izacard_leveraging_2021}.

In RAG models, a retrieval component first searches for and pulls in related documents, knowledge, or data from an external database or knowledge base. This information is then fed into a generative model, which uses the retrieved content to create a more informed and accurate response or output.

By incorporating RAG within an automated agent’s LLM, the agent can provide more relevant information, specific to its domain and use case~\cite{zhang_raft_2024,gao_retrieval-augmented_2024,jeong_study_2024}.

This is particularly relevant when considering the application of automated LLM agents to the help desk and professional consultation industries. Because these industries rely on relevant domain-specific topic knowledge, RAG can be an effective approach to address the shortcomings of broadly trained LLMs. This is significant, as LLM-powered automated agents appear to be well suited to the work required in these domains.

The help desk and professional consultation industries are broad and are present across almost all sectors of the economy. This includes customer support services, IT support services, administrative assistance, after-sales support and assistance, and several other types of support services. Although these service lines may appear to vary, they all rely on domain-specific knowledge, the ability to solve problems, and the ability to handle and mitigate errors to ensure that clients or customers have a good experience. Furthermore, they all rely on communication between a provider and a customer/client.

These are all tasks that automated LLM agents are designed to accomplish. However, their ability to do so effectively, particularly when compared to humans, remains unclear.

Answering these questions is significant, as, due to the applicability of automated LLM agents to help desk and professional consultation tasks, it stands to reason that there may be substantial impacts to the human workforce that already exists. In the US, almost 3 million people were employed as customer service representatives in 2023, earning a median hourly wage of \$19.08~\cite{noauthor_customer_nodate}. These positions may face potential redundancies as automation within the sector increases.

In 2021, the global help desk software market was valued at \$9.9 billion and is expected to grow at a compound annual growth rate of 9.4\%, reaching \$26.8 billion by 2032~\cite{saha_help_nodate}. Much of this growth is being driven by the demand for cloud-based solutions, which further stands to put existing jobs at risk.

Moreover, businesses in the US alone risk losing \$846 billion worth of sales because of poor customer service~\cite{fitzpatrick_customer_2024}, as studies have shown that 80\% of customers are likely to switch service providers after more than one bad experience with a brand~\cite{obrien_future_2024}.

Because of the high potential human and financial costs associated with the help desk, customer support, and professional consultation industries, it is imperative that the quality of assistance remains high, regardless of the extent to which automation becomes incorporated. Despite trends indicating that the use of AI technologies within these industries is growing~\cite{obrien_future_2024,tsymbal_ai_2024}, evidence suggests that clients/customers still prefer to get service from humans, particularly in nuanced or complex cases~\cite{tsymbal_ai_2024,zara_ai_2024}.

This all serves to highlight the growing need for a quantitative and comparative measure that can assess the performance of both humans and automated LLM agents in providing help desk and professional consultation assistance.

Benchmarking presents such a comparative tool. However, existing LLM benchmarks, as well as those targeting automated LLM agents, do not provide sufficient specific insight into the help desk and consultation industries (more information in Section~\ref{rw}). Given the potential impact automated LLM agents may have on the help desk and consultation industries, the development of a targeted benchmark specific to these industries is necessary to facilitate the integration of automated LLM agents into the industry. This gives rise to the primary research question of this study: \textbf{How can the performance of automated help desk and consultation agents be quantified and thus compared to the performance of human help desk and consultation agents?}

The current research presents \textbf{SelfScore}. This novel benchmark seeks to quantify the performance of automated LLM agents and humans when performing help desk and consultation tasks. By deriving its scores from intuitive and understandable math, the benchmark is more comprehensible and transparent than alternatives. This will help facilitate its adoption into domains wherein employees are less likely to have strong computer science or technical backgrounds.

For the current study, the research team developed automated help desk agents using current LLMs to assist with IT-related tasks to assess this newly developed benchmark. This gives rise to the second research question investigated herein: \textbf{How well do the developed automated LLM agents perform help desk tasks compared to humans?} Using the benchmark to quantify the performance of both humans and automated LLM agents, a direct comparison between the two can be drawn. This helps provide insight into the capabilities of current LLMs and the automated LLM agents that can be built upon them in order to assess the potential impact this technology might have on the future of the help desk and professional consultation industries.

\subsection{Related Work}\label{rw}
Currently, several benchmarks exist that aim to quantify the performance of LLMs. Some of these benchmarks focus on the ability of LLMs to engage in logical and critical thinking~\cite{williams_easy_2024,momennejad_evaluating_2023}. Others prioritize performance on general knowledge questions, ranging in difficulty from elementary school-level to expert-level~\cite{hendrycks_measuring_2021,rein_gpqa_2023}. These benchmarks are valuable for understanding the proficiencies of different LLMs on different tasks, and their observations have been considered in the development of SelfScore. For example, \cite{momennejad_evaluating_2023} notes that LLMs struggle with problems that rely on critical and logical thinking skills, while \cite{hendrycks_measuring_2021} observes that the largest LLMs are quite proficient in topic and domain knowledge.

However, these benchmarks do not target LLMs as a component of an automated agent. Where existing benchmarks do measure the performance of automated LLM agents~\cite{deng_mobile-bench_2024,wu_smartplay_2024,shridhar_alfred_2020,kinniment_evaluating_2024}, they focus on applications that are sufficiently different from the use case outlined in the current research to warrant the establishment of a new benchmark specific to the help desk and professional consultation industries. \cite{deng_mobile-bench_2024} assesses the capability for automated LLM agents to interact with mobile-based applications, while \cite{wu_smartplay_2024} focuses on automated LLM agents and their performance on games, \cite{shridhar_alfred_2020} assesses performance on visual-dependent household tasks, and \cite{kinniment_evaluating_2024} assesses the capacity of automated LLM agents to create software systems and self-replicate. Although the tasks measured in these benchmarks might share similar underlying dependencies (the ability to navigate applications and the internet, adaptability and critical thinking, and an understanding of computer science and systems, respectively), they are not specific enough to the tasks involved in help desk and professional consultation tasks to permit the assessment of automated LLM agents in this domain.

More closely related are benchmarks that consider “LLMs-as-Agents”~\cite{liu_agentbench_2023}. However, there remains a distinction between this use case and the use case outlined in the current study. Help desk, customer service, and professional consultation tasks may rely on the use of other services (text-to-speech/speech-to-text, internet browsing, code deployment, document generation, etc.) that fall outside the scope of what many LLMs alone are capable. This is an important aspect of SelfScore, as it evaluates the automated agent’s performance in its entirety, including that of other subsystems incorporated within the assessed autonomous LLM agent.

Furthermore, in many cases, these benchmarks evaluate performance on problems that have a well-defined correct answer~\cite{hendrycks_measuring_2021,rein_gpqa_2023,deng_mobile-bench_2024,shridhar_alfred_2020}. This is not always the case for help desk and professional consultation problems, where situational context and other nuances will impact a problem’s solution. Additionally, this suggests that they are not well suited to assess performance across a longer user-agent interaction, as they assess a single question-answer pair, which is unsuitable for the conversational nature of help desk and professional consultation processes. For benchmarks where this is not the case, their applicability to the help desk and professional industry remains an unresolved limitation for the research goals of the current study.
\section{SelfScore}\label{benchmark}
\subsection{Benchmark Criteria}
The proposed benchmark assesses an automated help desk agent’s ability to respond to a user question of varying complexity in a helpful manner. To accomplish this, one must first define and then calculate both the question’s complexity and the helpfulness of the agent’s response(s).

The benchmark also assesses the user’s interactions with the LLM agent, contextualized by the agent’s outputs. This ensures that especially effective or ineffective users do not artificially inflate or deflate the final score of a particular LLM agent.

\subsubsection{Complexity}\label{complexity}
The benchmark assesses the complexity of a human user’s query based on three primary criteria. Grading for these criteria is relative to the easiest and hardest problems in the corresponding domain. For the current study, this domain is computer and IT-related knowledge. However, the benchmark can apply to any LLM agent used for any help desk or professional consultation tasks. The benchmark is specifically tailored to help desk and consultation tasks in non-regulated domains. Though it was not designed with regulated domains in mind, it might also be applicable to these, as the extent of the impact of AI-enabled automation within regulated domains remains to be seen.

The criteria are as follows:

\begin{itemize}
    \item \textbf{Critical Thinking:} How much critical thinking does this problem require to solve.
    \item \textbf{Error Handling:} How likely is an error to occur while solving this problem, and, should an error occur, how significant will the impact of the error be and how difficult will it be to recover from it.
    \item \textbf{Topic Knowledge:} How much topic knowledge is required to solve the problem.
\end{itemize}

These criteria describe the three most significant independent factors that influence the complexity of a help desk interaction and are equally applicable regardless of the specific domain in which such an interaction occurs. Whether one uses an LLM agent to assist in IT contexts, customer support, or bureaucratic or administrative assistance, it is important to measure the level of necessary critical thinking, error handling, and topic knowledge to address user questions. It is important to note that these criteria are independent. For instance, certain problems might require in-depth topic knowledge, but their solutions might not require significant further critical thinking or elicit substantial errors. Take, for example, a programming problem wherein the rounding of a number does not work as expected. This error might result from the behavior of one of the programming language’s rounding functions. In this case, the solution might require the use of a different rounding function. Such a problem would require sufficient topic knowledge of the programming language but would not require further critical thinking or error management.

Furthermore, these criteria display continuous increases from simple help desk problems to challenging ones. As a user’s problem becomes more difficult, an LLM agent will require a higher level of at least one, if not more, of the identified criteria to solve it than would be necessary to solve an easier problem.

Moreover, LLMs – and, by extension, automated LLM agents – are suitable tools to address help desk and consultation problems and one can expect their performance on these criteria to improve proportionate to developments in the underlying technology. This makes these criteria suitable metrics by which to assess the overall performance of automated LLM agents for tasks of this type.

\subsubsection{Helpfulness}
The benchmark considers “helpfulness” at three separate points throughout the benchmarking process.

First, the benchmark considers the helpfulness of the initial input question. This refers to the level of information provided by the human user and how useful this information is in identifying and solving the user’s current problem. This assessment is necessary as the quality of the initial input question will have an impact on the ability of the LLM agent to address the underlying issue, regardless of how competent the agent itself is.

Next, the benchmark assesses the helpfulness of the user throughout the interaction. This consideration will rate how well a user is able to execute the agent’s instructions and will be continuous throughout the interaction between the user and LLM agent. This will primarily be determined by reviewing the user’s follow-up dialogue with the agent, following the initial question. As with the first helpfulness assessment, this consideration is necessary to ensure that user competency does not impact an agent’s final score.

Finally, the benchmark scores the helpfulness of the agent’s outputs. This consideration assesses whether the support agent could provide solutions relevant to the problem. This requires the agent’s outputs to be understandable, applicable, and constructive in addressing the user’s underlying problem.

The combination of these three separate “helpfulness” measures enables the benchmark to account for differences in user competency and ensures that the benchmark does not erroneously inflate or deflate scores depending on individual user interactions. These measures represent points during the help desk interaction where users can introduce variability. The benchmark must consider this to ensure it only accounts for the LLM agent’s abilities.

\subsubsection{Additional Considerations}
In addition to considerations regarding complexity and helpfulness, the benchmark can optionally assess other factors that may be relevant to benchmarking an LLM agent’s performance, depending on its specific use case.

For agents where operational costs may be a concern, benchmarking can optionally calculate and assess run costs. This permits assessors to weigh potential compromises between an agent’s performance using various foundational LLMs and any associated operational costs. In the current research, this consideration only applied to GPT-4 due to limitations of the deployed library LangChain.

For agents that incorporate text-to-speech (TTS) technology, the benchmark may also consider TTS-related aspects of the agent’s performance. These include the accuracy of the generated TTS, the comprehensibility of the TTS, and the naturalness of the TTS. These considerations are useful for agents that interact with users via audio interfaces, such as within call center applications, or similar.

\subsection{Benchmarking Process}\label{proc}
The process for benchmarking an LLM agent begins when the agent receives a question input as a problem. This input can originate from either a human user or a supplied testing dataset. At this stage, the benchmark calculates the weighted complexity score and assesses the helpfulness of the user’s initial question. Once the LLM agent responds to the initial question, the first “turn” of the interaction concludes.

The benchmark defines “turns” as an individual question-response pair within the user-agent interaction. The “interaction” refers to the complete set of turns that constitute the problem-solving or consultation task between the user and the agent. Turns loop until either the agent solves the problem or exceeds the maximum number of turns, at which point the interaction concludes. Assessors can set this number of turns according to their specific needs. For the current study, researchers set the maximum number of turns equal to 50.

Following the first turn, each subsequent turn involves the following steps:
\begin{enumerate}
    \item The benchmark assesses the user’s response to the agent for helpfulness. The agent then receives the user response.
    \item The agent generates its counter-response. The benchmark assesses this counter-response for LLM helpfulness. The user then receives the counter-response.
    \item The benchmark checks to determine if the agent has solved the user’s problem. In the affirmative case, the loop terminates, and the current turn ceases here. This check assesses the turn history and the latest LLM response to confirm that the user’s problem is, in fact, solved.
    \item If the assessor is assessing the per-turn run cost, the benchmark considers this here. There are three cases which control how this occurs.
    \begin{enumerate}
        \item Input and output tokens are the same cost. Calculate the total number of tokens used in this turn, multiplied by the token cost.
        \item Input and output tokens are priced differently. Calculate the number of input tokens multiplied by the input token cost and add to the number of output tokens multiplied by the output token cost.
        \item Stipulates a set per-turn cost, which is multiplied by the interaction’s total number of turns.
    \end{enumerate}
    \item The benchmark calculates the current turn’s quality. This score considers the helpfulness of the user responses and agent responses for this turn.
    \item The benchmark saves the user and LLM responses for this turn. These responses contribute to the assessment of overall helpfulness for both the user and agent across the entire interaction.
    \item The benchmark saves all scores, tokens, and other data for this turn for posterity (for instance, if there is a need to recalculate the benchmark scores for the current interaction).
    \item If the agent has not solved the user’s problem or has not exceeded the maximum number of turns, the loop restarts.
\end{enumerate}

Once the interaction between the user and the agent ends, either by way of solving the user’s question or by exceeding the maximum number of turns, the benchmark conducts calculations for final scoring. First, the benchmark calculates the average user and agent helpfulness across the entire interaction. Next, these scores determine the overall quality of the interaction. Finally, the benchmark calculates the interaction’s final score by comparing the problem’s weighted complexity with the average quality of the interaction.

The subsequent section (\ref{score} Scoring) provides the formulae for the calculations used throughout the benchmarking process.

\subsection{Scoring}\label{score}
Benchmark scoring combines quantitative and qualitative measures of the LLM agent’s performance. Consequently, subjectivity will play a role in scoring, as qualitative metrics related to problem complexity and response helpfulness are inherently subjective. The benchmark scores these subjective measures using an LLM. This permits the benchmark to efficiently assess on a larger scale than current alternatives. In all cases where the benchmark uses subjective measures, scores are based on a 10-point scale.

\subsubsection{Complexity Scoring}\label{comp_score}
As discussed in Section \ref{complexity}, the benchmark bases the complexity of a user’s initial question on three criteria: Critical Thinking, Error Handling, and Topic Knowledge, assessed relative to the easiest and hardest problems in the corresponding domain. The benchmark consolidates these three criteria into a single weighted complexity score, which aims to capture the overall complexity of the user’s problem.

The benchmark calculates the weighted complexity score using the following formula, which maintains the 10-point scale of its constituents:
\begin{multline}
    \text{Weighted Complexity} = (0.5 \times \text{Critical Thinking}) \nonumber \\
    + (0.4 \times \text{Error Handling}) + (0.1 \times \text{Topic Knowledge})
\end{multline}

Criterion weights consider the ability for LLMs to perform tasks related to the identified criteria to a high standard. As demonstrated in~\cite{hendrycks_measuring_2021}, LLMs are readily able to assist in tasks related to topic knowledge and domain knowledge transfer, thus it is not significantly considered in the weighted complexity score. However, topic knowledge remains an important aspect of addressing help desk and consultation questions and should factor into complexity considerations.

As demonstrated in~\cite{williams_easy_2024, momennejad_evaluating_2023}, LLMs currently struggle to demonstrate effective logical skills and critical thinking, as well as the ability to foresee and mitigate potential errors~\cite{kwong_long_2024,kamoi_evaluating_2024}. For this reason, these criteria are more important in determining the weighted complexity scores. Logical skills and critical thinking have been observed to be more challenging for LLMs than error handling, as demonstrated in comparing results from \cite{williams_easy_2024,kamoi_evaluating_2024}, hence their increased consideration within the weighted complexity score. This decision ensures that agents that are capable of these more complex tasks are more highly rewarded.

It is important to note that these weights are based on the conclusions of existing research into the capabilities of LLMs. They are not necessarily expected to be conclusive. More work is needed to finetune these weights to observe how weighted complexity might vary as these weights are adjusted. The above weights were used to generate the results of the current study and implementation.

\subsubsection{Helpfulness Scoring}
The benchmark independently scores the helpfulness of both the user and the agent every turn on a 10-point scale, where a 10 represents a perfectly helpful contribution to the ongoing interaction.

When assessing the user’s helpfulness, one should consider whether the user response in this turn demonstrates their ability to follow the previous turn’s instructions and communicate relevant new information clearly and concisely.

When assessing the agent’s helpfulness, one should consider whether the agent can provide solutions relevant to the current problem, considering the information provided by the user.

For the first turn of the interaction, the assessment of the user helpfulness differs slightly. This distinction is necessary because this turn involves the initial question. As such, there are no previous instructions for the user to follow. Consequently, one should instead consider how helpful this initial question is based on how much potentially helpful information it contains, relevant to solving the problem.

Once the interaction has concluded, the benchmark calculates the overall average helpfulness scores for both the user and the agent using the following formulae:
$$\text{Average User Helpfulness} = \frac{\sum \text{Per Turn User Helpfulness Scores}}{\text{Number of Turns}}$$

$$\text{Average LLM Helpfulness} = \frac{\sum \text{Per Turn LLM Helpfulness Scores}}{\text{Number of Turns}}$$

\subsubsection{Quality Scoring}
The quality score arises from the helpfulness ratings and describes the overall quality of the interactions. The benchmark uses the following formulae:
$$\text{Average Quality} = \frac{\text{Average LLM Helpfulness}}{\text{Average Human Helpfulness}}$$

Quality score ensures that the benchmark does not penalize effective agents due to ineffective users, and that effective users do not artificially inflate the score of poor agents.

\subsubsection{Final Scoring}
The final benchmark score considers both the weighted complexity score, derived from the user’s initial question, as well as the average quality of the entire interaction, using the following formula:
$$\text{Final Score} = \frac{\text{Weighted Complexity} + \text{Average Quality}}{2} \times 10$$

The final score provides a theoretical maximum score of 100. This would require both the human user and the LLM agent to be perfectly helpful over the entire interaction on a maximally complex problem in terms of critical thinking, error management, and topic knowledge.

In practice, it is likely difficult for an agent to achieve such a score. However, as agent design and the capabilities of foundational models improve, it seems likely that scores will improve in a continuous manner as agents are able to solve more complex problems and communicate their solutions more effectively to users.
\section{Current Implementation}\label{implementation}
To assess and validate the proposed benchmark, the research team designed and implemented automated LLM agents to assist in IT-domain help desk tasks. These agents use GPT-4, Mixtral7b, and Mixtral8x7b as foundational models. 

The agents received tasks originating from a dataset comprised of Stack Exchange forum posts. The dataset is derived from an anonymized archive dump of all user-contributed content on Stack Exchange, which included posts, users, votes, comments, badges, tags, post history, and post links~\cite{stack_exchange_inc_stack_nodate}. Posts were selected from the archive dump based first on whether they had an accepted answer. Following this, further selection was based on the number of answers’ upvotes to ensure that only high-quality postings were included in the dataset. From this selection, the text content and title of the post, the response, and the number of upvotes were considered relevant going forward.

For the current implementation of the benchmark, the research team used LLMs to assist in a variety of tasks. These tasks include performing large-batch information extraction, assessments of problem complexity, and assessments of user and agent helpfulness, among others. Where applicable, relevant prompts accompany descriptions of LLM use.

\subsection{Data Pre-Processing and Initial Calculations}
First, researchers converted the top entries from the Stack Exchange dataset from XML to JSON format for ease of use and consistency. Selection for the dataset’s top entries considered the number of upvotes on a forum posting’s answer. This threshold varied depending on whether the agent used RAG or not. For agents that were not using RAG, researchers set the minimum answer upvote to 100, which resulted in a pool of 1,164 entries. For agents using RAG, this threshold was lowered such that the pool was approximately twice as large (total of 2,360 entries). From this pool, random selection determined which entries were used for RAG and which were used for benchmark testing and validation, ensuring a 50/50 split to maintain validation pool size consistency between runs.

Since user questions were occasionally lengthy, an LLM extracted a summarized version of the question and the underlying problem from the posting’s solution.

The prompt for question extraction follows:
\texttt{Dumb this question down and summarize it in one or two sentence(s):<Ques-\allowbreak{}tion>.}

Originally, the question extraction prompt used more formal vocabulary (e.g., “simplify this question,” “give a concise version of this question”). However, this version provided the best results.

The prompt for underlying problem extraction follows: 
\texttt{Extract a problem statement from this post. For example, "The computer is not plugged in", or "The DNS servers are down". Respond with only the problem: <Problem>.}

This approach minimizes the level of superfluous details provided to the LLM agent and minimizes required input tokens, where necessary. Furthermore, LLMs are well suited to information extraction tasks, as demonstrated in \cite{xu_large_2024}.

During runtime, the calculation for weighted complexity score considers the summarized version of the question, while the solution check (Point 3 in Section ~\ref{proc} Benchmarking Process) uses the extracted underlying problem.

To ensure the qualitative assessments are as representative as possible, multiple LLMs perform the assessments for complexity and helpfulness across different testing runs. For example, when using Mixtral8x7b as a foundational model for the agent, one testing run uses GPT-4 to perform qualitative assessment, while another uses Mixtral8x7b, and so on. This permits the direct comparisons of final scores, quality, and complexity and helpfulness assessments to confirm that LLM information extraction and qualitative assessment is robust and suitable for the current task.

\subsection{Agent Design}
\begin{figure} [ht]
    \begin{center}  
        \includegraphics[width=3.3in]{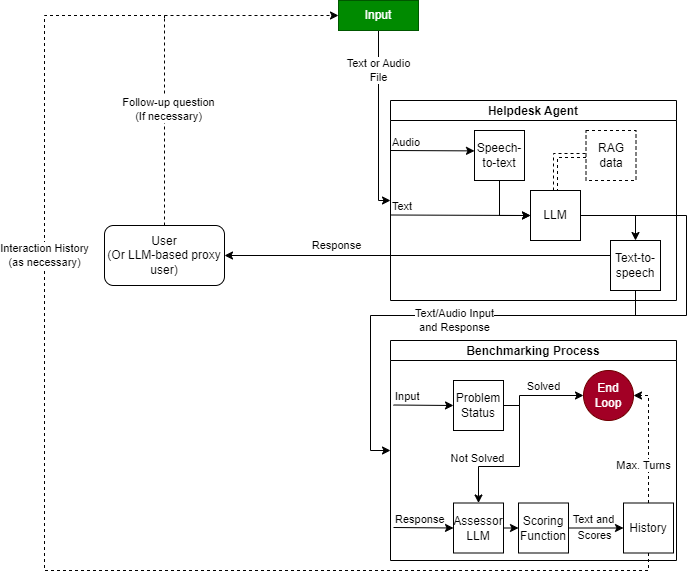}
        \caption{\small \sl Interaction Flow Scheme.
        \label{fig:benchmark_diagram}}  
    \end{center}
\end{figure}

The help desk agents share a similar design, differing only in the foundational LLM used for language processing and generation. Figure \ref{fig:benchmark_diagram} visualizes the flow between the agent, the user, and the benchmark throughout the interaction.

The agent begins by receiving an input. This input is either the initial problem question or a follow-up question. For the current implementation, the source for these inputs is either a dataset entry (used to provide a human baseline or control assessment) or an LLM generated response which acts as a proxy for the agent’s user. This approach facilitates large-scale assessment. In cases where the inputs are audio files, OpenAI’s Whisper speech-to-text~\cite{noauthor_introducing_2022} converts the audio file to a string. Otherwise, the agent receives the input as text directly.

Once the agent receives the input, it passes this input, along with any interaction history, to the inference mechanism of the foundational LLM's backend. The LLM then processes the input to generate a text-based response, which is subsequently output back to the agent. Where relevant, the agent also returns the input and output tokens used.

A script then cleans the LLM’s text output to remove unsupported characters by the agent’s TTS mechanism. This step is necessary due to the specific TTS component~\cite{eren_coqui_2021} that the current agent implementation uses. The agent then writes the audio file and returns its response as both audio and text. At this point, the agent also returns the input to the benchmark to preserve interaction history.

It is important to note that, in the current implementation, the benchmark manages the interaction’s history as opposed to the agent itself. This approach ensures that the benchmark can enforce the maximum number of turns per interaction and can determine when the agent has solved the user’s problem. Currently, the agent itself is not able to reset interaction history on its own. This task is managed by the benchmark, which passes history to the agent, as necessary. In another implementation, the agent could exclusively manage history on an interaction-by-interaction basis.

Finally, the turn concludes, and the agent receives a new input from the dataset or from an LLM generated user proxy. The agent then repeats this process until it reaches the maximum number of turns or until it solves the problem.

In the case that an agent does not use RAG, it relies on the following system prompt: 
\texttt{A user is having a problem. Respond with simple and helpful instructions most likely to guide the user to a solution. Only provide one solution at a time. Never give instructions to contact external or professional services. Never suggest contacting external or professional services.}

Although the last sentence of this prompt is a repeated order, the research team identified that this minimized the chance that the output would suggest contacting external or professional services.

In the case that an agent does use RAG, there is no provided system prompt. Instead, a system prompt is either unnecessary and inferred by RAG data, or the developer explicitly defines it in the agent’s backend.

\subsection{Results}
The average final scores for all assessment runs of the benchmark are in the table below.
\begin{table}[h]
    \centering
    \caption{Average Final Scores of Agents}
    \begin{tabular}{ll}
        \toprule
        \textbf{Agent LLM} & \textbf{Average Final Score} \\
        \midrule
        GPT-4\tablefootnote{GPT-4 version: gpt-4-1106-preview. This model is used throughout the current research and is referred to simply as GPT-4.} + RAG & 29.35 \\
        Mixtral 8x7b + RAG & 28.79 \\
        Mistral 7b + RAG & 28.46 \\
        Mistral 7b & 28.08 \\
        Mixtral 8x7b & 27.70 \\
        Mixtral 8x7b\tablefootnote{This assessment run of the Mixtral 8x7b agent used GPT-4 for benchmarking tasks, while the preceding run used Mixtral 8x7b for all tasks.} & 23.65 \\
        GPT-4 & 23.40 \\
        Human & 23.12 \\
        \bottomrule
    \end{tabular}
\end{table}

The results shown in the table indicate that agents incorporating RAG performed better than those that did not. The top 3 performing automated LLM agents all achieved similar average final scores, with a difference of only 0.89 between the top performer and third performer’s average final scores. An ANOVA indicated that the difference in final scores for the top 3 performers is not statistically significant ($F=2.00,p=0.135$).

However, when considering all assessed agents’ final scores, an ANOVA revealed that the observed difference was statistically significant ($F=84.7,p<0.001$). Appendix \ref{app} contains post-hoc test results. These results suggest that the benchmark can effectively discern between differently capable LLMs. Furthermore, they reinforce the observation that agents incorporating RAG are better suited for domain-specific assistance.

The detailed results for the top three performing agents, as well as those from the human baseline, are found below. Pairwise comparisons using Mann-Whitney U tests accompany the result descriptions to determine statistical significance of between group results, where necessary. The Mann-Whitney U statistical test accounts for the non-normal distribution observed in key metric scoring.

Each result name is split into three parts, divided by underscores. The first part refers to the foundational LLM used for the agent, the next part refers to the model used to generate the dataset's complexity scores, and the third part refers to the model used for evaluation of responses during benchmarking. For instance, the results \texttt{mixtral-8x7b\_gpt-4-1106-preview\_gpt-4-1106-preview} uses Mixtral 8x7b as the agent's foundational LLM, uses GPT-4-1106 preview for the dataset complexity generation, and uses GPT-4-1106 preview for the response evaluation. 

Note that the exception to this is \texttt{human\allowbreak{}\_gpt\allowbreak{}-4\allowbreak{}-1106\allowbreak{}-preview\allowbreak{}\_gpt\allowbreak{}-4\allowbreak{}-1106\allowbreak{}-preview}, which represents the current control. These results do not use an LLM agent to generate responses, and instead use the existing human interactions derived from the Stack Exchange forum. This provides a baseline against which to compare the performance of automated LLM agents. In this case, GPT-4 performed benchmarking tasks by assessing complexity and helpfulness. The detailed plots of key insights follow this naming scheme.

\begin{figure} [ht]
    \begin{center}  
        \includegraphics[width=3.3in]{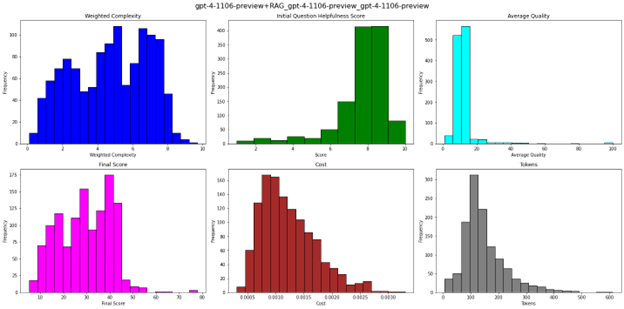}
        \caption{\small \sl GPT-4 + RAG Detailed Results.
        \label{fig:gpt-4_results}}  
    \end{center}
\end{figure}

Figure ~\ref{fig:gpt-4_results} displays the detailed results of the top performing LLM agent. This agent uses GPT-4 as a foundational model and incorporates RAG. GPT-4 was also used to perform subjective assessments regarding complexity and helpfulness, as well as simulate the responses of the agent’s user. Ratings for complexity scores demonstrate a multimodal distribution, with few complexity scores near the extremes. Helpfulness scores for this assessment run are, in general, highly rated. However, average interaction quality scores are generally low. This results in a distribution of final scores that are mostly in the lower half of the possible score range. This distribution of final scores is similar across all assessment runs, however this agent demonstrated higher outlier final scores in the 60s and 70s.

\begin{figure} [ht]
    \begin{center}  
        \includegraphics[width=3.3in]{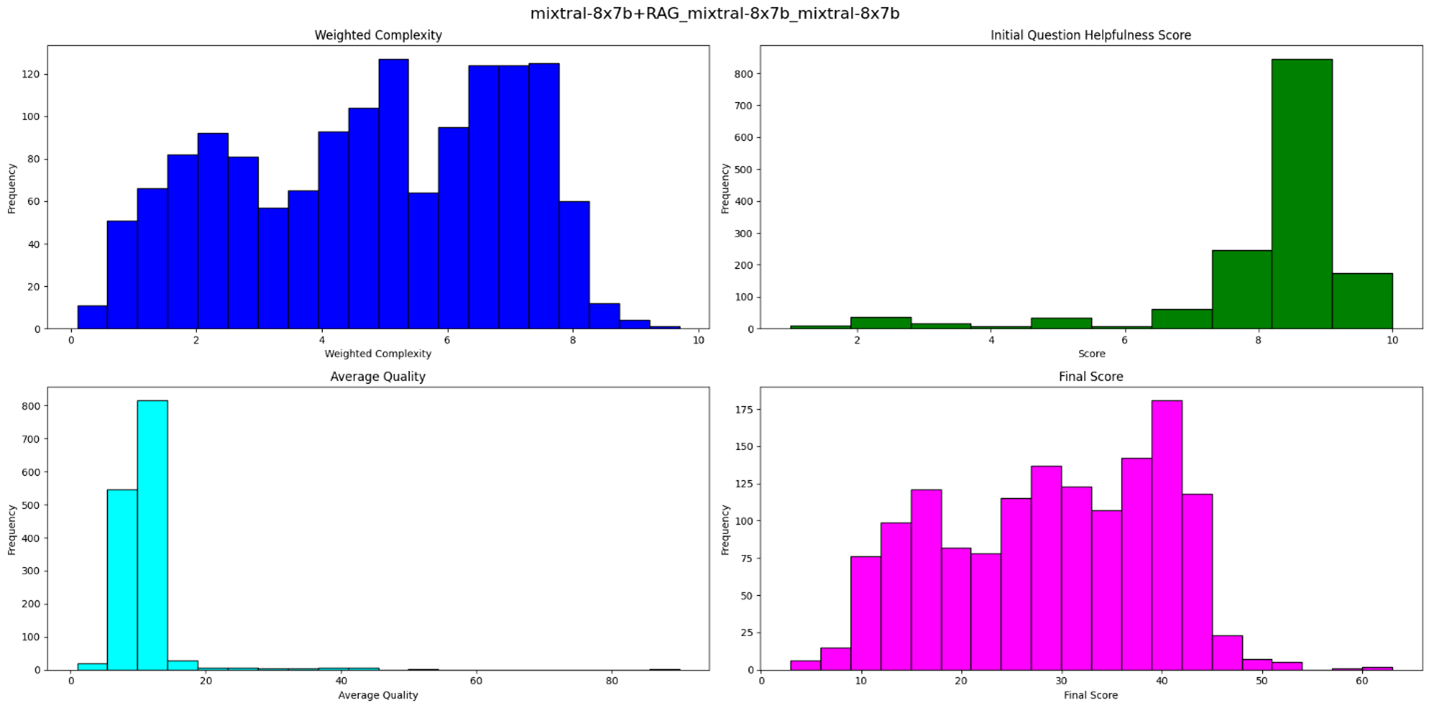}
        \caption{\small \sl Mixtral 8x7b + RAG Detailed Results.
        \label{fig:mixtral8x7b+RAG_results}}  
    \end{center}
\end{figure}

Figure \ref{fig:mixtral8x7b+RAG_results} shows the results for the Mixtral 8x7b-based model. These results are generally similar to those displayed in Figure \ref{fig:gpt-4_results}. Notably, helpfulness scores were rated higher than in the assessment run depicted in Figure \ref{fig:mixtral8x7b+RAG_results}, however average interaction quality and final scores were lower, with fewer high score outliers. The bulk of final scores from both assessment runs show a similar distribution, though there were no final scores above 70.

\begin{figure} [ht]
    \begin{center}  
        \includegraphics[width=3.3in]{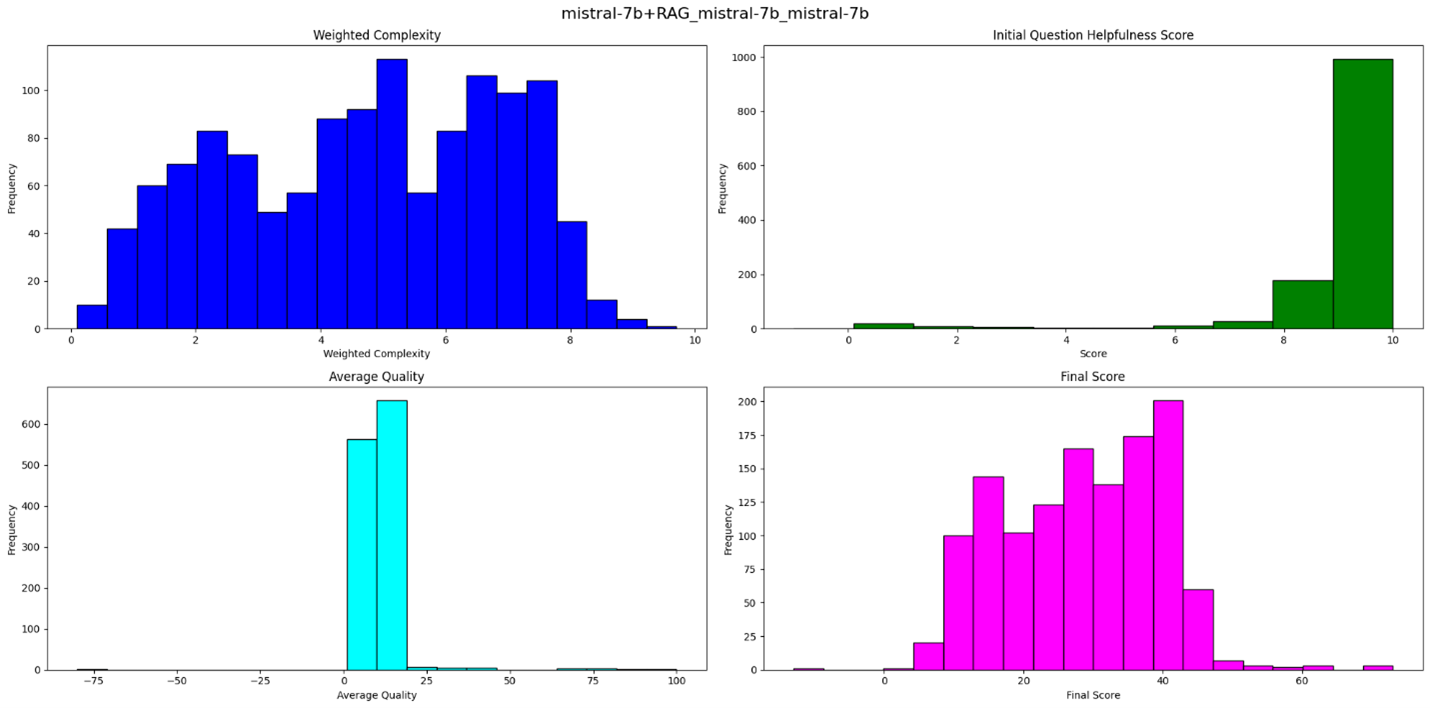}
        \caption{\small \sl Mistral 7b + RAG Detailed Results.
        \label{fig:mistral7b+RAG_results}}  
    \end{center}
\end{figure}

The agent based on Mistral 7b (Figure \ref{fig:mistral7b+RAG_results}) again displays similar trends to those observed in Figures \ref{fig:gpt-4_results} and \ref{fig:mixtral8x7b+RAG_results}. Helpfulness scores are higher than the two preceding results.

\begin{figure} [ht]
    \begin{center}  
        \includegraphics[width=3.3in]{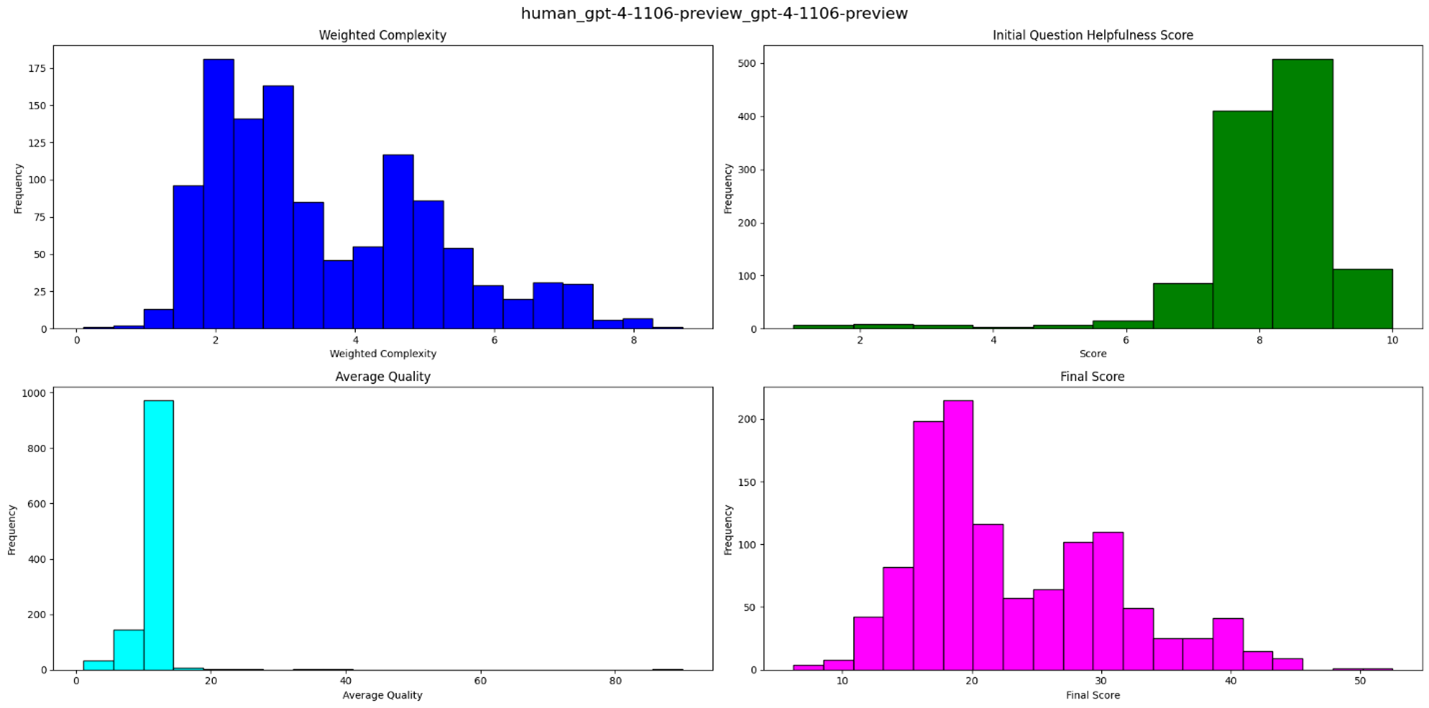}
        \caption{\small \sl Human Baseline Detailed Results.
        \label{fig:human_results}}  
    \end{center}
\end{figure}

The human baseline results are displayed in Figure \ref{fig:human_results} These results are derived directly from the human interactions from the original Stack Exchange dataset. GPT-4 performed benchmarking tasks regarding complexity and helpfulness assessment. Not only were helpfulness scores significantly lower ($U=4.33e^6,p<0.001$), but the benchmark scored the human interactions as significantly worse than all tested agents ($U=4.09e^6,p<0.001$).

\begin{figure} [ht]
    \begin{center}  
        \includegraphics[width=3.3in]{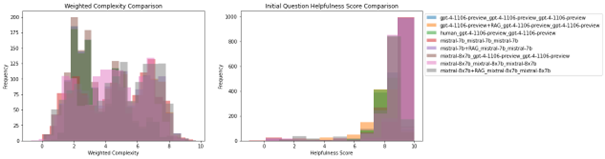}
        \caption{\small \sl Comparison of Complexity and Helpfulness Assessments.
        \label{fig:comparison}}  
    \end{center}
\end{figure}

Figure \ref{fig:comparison} displays comparisons of the subjective assessments performed by the LLMs to assess the feasibility of using LLMs for these types of assessments.

These comparisons display a statistically significant difference within the subjective assessments (Complexity: $F=95.4,p<0.001$; Helpfulness: $F=76.1,p<0.001$). This result is unsurprising, as training data differs between LLMs. Thus, depending on the content of this training data, LLMs will have differing levels of competency for different domains. The final choice for which LLM to use for benchmarking assessments will eventually depend on the specific domain and use case.
\section{Discussion}\label{disc}
\subsection{Contributions}
The proposed benchmark represents a significant contribution to LLM agent benchmarking by permitting large scale testing of help desk and consultation agents. By using LLMs to conduct the benchmark’s required assessments, significantly more data can be considered during scoring due to the rapidity with which LLMs can process written text. Thus, assessors can determine the performance of automated LLM agents across more numerous and more varied interactions than previous benchmarks allowed.

Furthermore, the scoring is simple and intuitive. Final scores are based on a logical 100-point scale and are derived from simple mean calculations. Constituent criteria are all scored on an intuitive 10-point scale based on clear definitions that are comprehensible to non-expert analysts. This approach broadens the benchmark’s potential userbase and accessibility, thus facilitating its widespread adoption to a plethora of help desk and professional consultation industries, where existing human staff might not be proficient in AI and LLM technologies or assessment tools.

The current implementation, including the developed automated LLM agents, builds on this contribution by reinforcing the observation that RAG provides significant benefits to help desk-type tasks. Domain-specific RAG material led to significant improvements in the performance of automated LLM agents and their ability to helpfully solve user problems. This result reinforces the observations of \cite{zhang_raft_2024,gao_retrieval-augmented_2024, jeong_study_2024} and provides further evidence for the benefits of sourcing relevant additional material to use for RAG when designing this type of LLM agent.

Additionally, the current study’s approach of simulating the user side of the interaction facilitates the generation of test sets. This permits LLM agent developers to simulate interactions and assess the performance of their developed agents prior to release.

Based on the results of the current study, automated LLM agents are already capable of solving forum-based IT-related problems better than humans. Although all forum postings in the human baseline were marked as solved, the benchmark scores indicated that the human baseline was still the worst performing cohort of those assessed in the current study. These results indicate a pressing need to investigate the social ramifications of highly capable automated LLM agents on the status quo of the help desk and professional consultation industries.

\subsection{Limitations}\label{limits}
One might argue that a limitation of the current benchmark is its reliance on LLMs for subjective judgments. There is some validity to this criticism, since, as demonstrated in Figure \ref{fig:human_results}, different LLMs will differ in their scoring. Moreover, a single LLM might assign different scores to the same question or response across multiple runs.

However, given the subjective nature of these judgments, it is reasonable to expect that human experts may also reach different conclusions on the complexity and helpfulness scoring for the same interaction. Additionally, human experts are subject to errors arising due to task exhaustion, lack of concentration, or task-related bias. For instance, experts might misjudge the complexity of problems for non-experts due to their familiarity with the domain. Using LLMs for such tasks will help minimize these types of errors. This point is particularly salient considering the current approach’s ability to efficiently test a large batch of user-agent interactions.

This limitation might be partially mitigated by explicitly formalizing a rubric and providing examples for each point on the rating scale for each assessment criterion.

\subsection{Future Work}
The logical next step in further assessing the current benchmarking approach is to compare the subjective assessments made by LLMs to those made by relevant human domain experts. This will provide a baseline against which to determine the efficacy of using LLMs for the benchmark’s constituent subjective assessments. Results will demonstrate the extent to which assessments made by LLMs align with those made by human domain experts. Should results indicate divergences between human and LLM scoring, these can then be mitigated by the data generated from the future work itself. For instance, this data can be used for RAG or to otherwise fine-tune the LLMs used in benchmarking assessments. This work will need to be done on a domain-by-domain basis but could help the assessing LLMs make conclusions that are more in line with human assessors, should there be divergences.

Further future work could focus on the collection and benchmarking of live human-agent interactions. This work would differ from the current implementation in that user problems and the subsequent interactions would not be derived from a dataset of forum postings. Instead, live users would interact with an agent in real-time. These crowd-sourced interactions can then be used to assess ready-to-be-deployed agents. This can help provide insight into how the benchmark performs across different points in an agent’s development and deployment lifecycle. Moreover, this type of work can focus on collecting data regarding the performance of agents in sectors outside of IT, thus investigating the effectiveness of the benchmark across different domains. 

Outside of internet forums, help desk and consultation interactions often occur over the phone or in person. In these cases, the capacity for agents to respond with a high degree of humanness in real-time will be more relevant. Additionally, interactions might be more open-ended than those in online forums. Although the current implementation did not include these types of interactions, the benchmark is designed to account for them. Consequently, data of this nature will help further assess the benchmark.

In addition to the improvements that can be gained through the comparison of human/LLM assessments and broader testing data, the benchmark might benefit from explicit grading rubrics and point-by-point examples. As discussed in \ref{limits} Limitations, establishing a clear description and example for each criterion point for the helpfulness and complexity assessments may enable different LLMs to reach more consistent conclusions on these subjective measures, both across scoring runs and between different models.

Similarly (as briefly discussed in Section \ref{comp_score} Complexity Scoring), more work needs to be done to finetune and adjust the weights within the weighted complexity function. The current weights consider what is hard for LLMs and reward those that can better accomplish these difficult tasks. However, it is unclear whether these weights are optimized or whether there might be alternative weights that better describe overall problem complexity. For example, a possible adjustment to the weights could consider what is hard for humans. In this case, topic knowledge would likely be weighted more heavily than critical thinking, as humans do not struggle with logical thinking like LLMs.

Finally, the results of the study indicate the need to investigate the potential impact of increased and highly capable automation on the existing help desk and consultation industries, with a specific focus on the social impact. The results of the current study suggest that workers within the help desk and consultation might be at risk of job replacement. This type of research could involve longitudinal studies into the increasing prevalence of AI technologies within the industry, how these trends have impacted jobs over the last few years, and the development and analysis of predictive models based on this data. Furthermore, information about what other roles human help desk workers fill (documentation, reporting, miscellaneous administrative tasks, etc.), the relative costs of humans to AI agents (including their development and deployment), and other considerations will be necessary to reach conclusions about the long-term impact of AI on this sector. The current benchmark provides a quantitative measurement to facilitate this work, which will be instrumental in conducting these follow-up investigations.
\section{Conclusions}
Based on the current results, automated LLM agents are becoming more capable than humans within the investigated subsection of the help desk and professional consultation domain. Should this trend continue, and should AI technologies continue to improve and proliferate, one can expect that jobs in this industry may be heavily impacted. This is significant as the help desk industry (specifically the customer service sector) accounts for 3 million jobs in the US alone. Considering the global help desk software market is already valued at \$9.9 billion and is growing at a considerable rate, this impact will be felt worldwide.

These results, in addition to trends indicating that AI is already playing an increasingly large role in the industry, suggest that an impending transitionary period is approaching. Due to the potential human and financial impact of this transition, there exists a collective moral obligation to ensure that it can occur in such a way as to minimize potential harm and to uphold high service standards.

To ensure that this is carried out in such a manner, benchmarks, such as SelfScore, can be leveraged to ensure that newly implemented technologies provide added value to their human counterparts. Due to the importance of high-quality service, it is ill-advised to replace human workers should they perform better than automated alternatives. However, in the contrary case, it is necessary to assess the implications of industry redundancies and the associated broader social impacts. It may prove beneficial to investigate alternative solutions, such as the deployment of automated technologies in a collaborative role with existing human workforces. Final assessment will need to be performed on a case-by-case basis, given the differing capabilities and knowledgebase of LLMs and automated LLM agents.

SelfScore will enable these assessments and provide intuitive and easily comprehensible results, which is of considerable value given the complexity of LLMs and the persistence of the “black box” phenomenon. Furthermore, SelfScore will help facilitate future research into the broader societal and socioeconomic impacts of an ongoing transition towards automation within the help desk and consultation industries.

In developing SelfScore, the research team hopes that the integration of automation technologies can be carried out in a way that is considerate of the potential human and social ramifications, both within the help desk and professional consultation industries and more broadly. Such considerations should remain paramount in future studies on AI and its deployment, ensuring that technological advancements align with ethical standards and societal well-being.

\bibliographystyle{ACM-Reference-Format}
\bibliography{references.bib}


\begin{thebibliography}{48}


\ifx \showCODEN    \undefined \def \showCODEN     #1{\unskip}     \fi
\ifx \showDOI      \undefined \def \showDOI       #1{#1}\fi
\ifx \showISBNx    \undefined \def \showISBNx     #1{\unskip}     \fi
\ifx \showISBNxiii \undefined \def \showISBNxiii  #1{\unskip}     \fi
\ifx \showISSN     \undefined \def \showISSN      #1{\unskip}     \fi
\ifx \showLCCN     \undefined \def \showLCCN      #1{\unskip}     \fi
\ifx \shownote     \undefined \def \shownote      #1{#1}          \fi
\ifx \showarticletitle \undefined \def \showarticletitle #1{#1}   \fi
\ifx \showURL      \undefined \def \showURL       {\relax}        \fi
\providecommand\bibfield[2]{#2}
\providecommand\bibinfo[2]{#2}
\providecommand\natexlab[1]{#1}
\providecommand\showeprint[2][]{arXiv:#2}

\bibitem[noa(2022)]%
        {noauthor_introducing_2022}
 \bibinfo{year}{2022}\natexlab{}.
\newblock \bibinfo{title}{Introducing {Whisper}}.
\newblock
\newblock
\urldef\tempurl%
\url{https://openai.com/index/whisper/}
\showURL{%
\tempurl}


\bibitem[noa(2023a)]%
        {noauthor_meta_2023}
 \bibinfo{year}{2023}\natexlab{a}.
\newblock \showarticletitle{Meta prepares {AI}-powered chatbots in attempt to retain users, {Financial} {Times} reports}.
\newblock \bibinfo{journal}{\emph{Reuters}} (\bibinfo{date}{Aug.} \bibinfo{year}{2023}).
\newblock
\urldef\tempurl%
\url{https://www.reuters.com/technology/meta-prepares-ai-powered-chatbots-attempt-retain-users-ft-2023-08-01/}
\showURL{%
\tempurl}


\bibitem[noa(2023b)]%
        {noauthor_musk_2023}
 \bibinfo{year}{2023}\natexlab{b}.
\newblock \showarticletitle{Musk says his {AI} firm {xAI} is rolling out chatbot {Grok} to {X} {Premium}+ subscribers}.
\newblock \bibinfo{journal}{\emph{Reuters}} (\bibinfo{date}{Dec.} \bibinfo{year}{2023}).
\newblock
\urldef\tempurl%
\url{https://www.reuters.com/technology/musk-says-his-ai-firm-xai-is-rolling-out-chatbot-grok-x-premium-subscribers-2023-12-07/}
\showURL{%
\tempurl}


\bibitem[noa(2023c)]%
        {noauthor_what_2023-1}
 \bibinfo{year}{2023}\natexlab{c}.
\newblock \bibinfo{title}{What {Are} {Large} {Language} {Models} ({LLMs})?}
\newblock
\newblock
\urldef\tempurl%
\url{https://www.ibm.com/topics/large-language-models}
\showURL{%
\tempurl}


\bibitem[noa(2023d)]%
        {noauthor_what_2023}
 \bibinfo{year}{2023}\natexlab{d}.
\newblock \bibinfo{title}{What {Is} {Enterprise} {AI}?}
\newblock
\newblock
\urldef\tempurl%
\url{https://www.ibm.com/topics/enterprise-ai}
\showURL{%
\tempurl}


\bibitem[noa(2024a)]%
        {noauthor_ai_nodate}
 \bibinfo{year}{2024}\natexlab{a}.
\newblock \bibinfo{title}{{AI} in {Finance}: {Applications}, {Examples} \& {Benefits}}.
\newblock
\newblock
\urldef\tempurl%
\url{https://cloud.google.com/discover/finance-ai}
\showURL{%
\tempurl}


\bibitem[noa(2024b)]%
        {noauthor_chatgptcom_nodate}
 \bibinfo{year}{2024}\natexlab{b}.
\newblock \bibinfo{title}{chatgpt.com {Traffic} {Analytics}, {Ranking} \& {Audience} [{August} 2024]}.
\newblock
\newblock
\urldef\tempurl%
\url{https://www.similarweb.com/website/chatgpt.com/}
\showURL{%
\tempurl}


\bibitem[noa(2024c)]%
        {noauthor_chatopenaicom_nodate}
 \bibinfo{year}{2024}\natexlab{c}.
\newblock \bibinfo{title}{chat.openai.com {Traffic} {Analytics}, {Ranking} \& {Audience} [{August} 2024]}.
\newblock
\newblock
\urldef\tempurl%
\url{https://www.similarweb.com/website/chat.openai.com/}
\showURL{%
\tempurl}


\bibitem[noa(2024d)]%
        {noauthor_customer_nodate}
 \bibinfo{year}{2024}\natexlab{d}.
\newblock \bibinfo{title}{Customer {Service} {Representatives}}.
\newblock
\newblock
\urldef\tempurl%
\url{https://www.bls.gov/ooh/office-and-administrative-support/customer-service-representatives.htm}
\showURL{%
\tempurl}


\bibitem[noa(2024e)]%
        {noauthor_introduction_nodate}
 \bibinfo{year}{2024}\natexlab{e}.
\newblock \bibinfo{title}{Introduction to {Large} {Language} {Models} {\textbar} {Machine} {Learning}}.
\newblock
\newblock
\urldef\tempurl%
\url{https://developers.google.com/machine-learning/resources/intro-llms}
\showURL{%
\tempurl}


\bibitem[noa(2024f)]%
        {noauthor_what_nodate}
 \bibinfo{year}{2024}\natexlab{f}.
\newblock \bibinfo{title}{What is enterprise {AI}?}
\newblock
\newblock
\urldef\tempurl%
\url{https://cloud.google.com/discover/what-is-enterprise-ai}
\showURL{%
\tempurl}


\bibitem[Breunig et~al\mbox{.}(2017)]%
        {breunig_building_nodate}
\bibfield{author}{\bibinfo{person}{Matthias Breunig}, \bibinfo{person}{Matthias Kässer}, \bibinfo{person}{Heinz Klein}, {and} \bibinfo{person}{Jan~Paul Stein}.} \bibinfo{year}{2017}\natexlab{}.
\newblock \showarticletitle{Building smarter cars with smarter factories: {How} {AI} will change the auto business}.
\newblock  (\bibinfo{year}{2017}).
\newblock


\bibitem[Buchanan et~al\mbox{.}(2017)]%
        {buchanan_digital_nodate}
\bibfield{author}{\bibinfo{person}{Jennifer Buchanan}, \bibinfo{person}{Beth Kelley}, {and} \bibinfo{person}{Alicia Hatch}.} \bibinfo{year}{2017}\natexlab{}.
\newblock \showarticletitle{Digital workplace and culture {How} digital technologies are changing the workforce and how enterprises can adapt and evolve}.
\newblock  (\bibinfo{year}{2017}).
\newblock


\bibitem[Cartwright(2023)]%
        {cartwright_agriculture_nodate}
\bibfield{author}{\bibinfo{person}{Mark Cartwright}.} \bibinfo{year}{2023}\natexlab{}.
\newblock \bibinfo{title}{Agriculture in the {British} {Industrial} {Revolution}}.
\newblock
\newblock
\urldef\tempurl%
\url{https://www.worldhistory.org/article/2191/agriculture-in-the-british-industrial-revolution/}
\showURL{%
\tempurl}


\bibitem[Deng et~al\mbox{.}(2024)]%
        {deng_mobile-bench_2024}
\bibfield{author}{\bibinfo{person}{Shihan Deng}, \bibinfo{person}{Weikai Xu}, \bibinfo{person}{Hongda Sun}, \bibinfo{person}{Wei Liu}, \bibinfo{person}{Tao Tan}, \bibinfo{person}{Jianfeng Liu}, \bibinfo{person}{Ang Li}, \bibinfo{person}{Jian Luan}, \bibinfo{person}{Bin Wang}, \bibinfo{person}{Rui Yan}, {and} \bibinfo{person}{Shuo Shang}.} \bibinfo{year}{2024}\natexlab{}.
\newblock \bibinfo{title}{Mobile-{Bench}: {An} {Evaluation} {Benchmark} for {LLM}-based {Mobile} {Agents}}.
\newblock
\newblock
\urldef\tempurl%
\url{https://doi.org/10.48550/arXiv.2407.00993}
\showDOI{\tempurl}
\newblock
\shownote{arXiv:2407.00993 [cs]}.


\bibitem[Early(2023)]%
        {early_can_2023}
\bibfield{author}{\bibinfo{person}{Catherine Early}.} \bibinfo{year}{2023}\natexlab{}.
\newblock \showarticletitle{Can artificial intelligence help wean city-dwellers from their cars?}
\newblock \bibinfo{journal}{\emph{Reuters}} (\bibinfo{date}{Dec.} \bibinfo{year}{2023}).
\newblock
\urldef\tempurl%
\url{https://www.reuters.com/sustainability/climate-energy/can-artificial-intelligence-help-wean-city-dwellers-their-cars-2023-12-18/}
\showURL{%
\tempurl}


\bibitem[Eren and {The Coqui TTS Team}(2021)]%
        {eren_coqui_2021}
\bibfield{author}{\bibinfo{person}{Gölge Eren} {and} \bibinfo{person}{{The Coqui TTS Team}}.} \bibinfo{year}{2021}\natexlab{}.
\newblock \bibinfo{title}{Coqui {TTS}}.
\newblock
\newblock
\urldef\tempurl%
\url{https://doi.org/10.5281/zenodo.6334862}
\showDOI{\tempurl}


\bibitem[Fitzpatrick(2024)]%
        {fitzpatrick_customer_2024}
\bibfield{author}{\bibinfo{person}{Klarissa Fitzpatrick}.} \bibinfo{year}{2024}\natexlab{}.
\newblock \bibinfo{title}{Customer {Service} {Statistics}: 2024 {Report}}.
\newblock
\newblock
\urldef\tempurl%
\url{https://www.ringover.com/blog/customer-service-statistics}
\showURL{%
\tempurl}


\bibitem[Gao and Feng(2023)]%
        {gao_ai-driven_2023}
\bibfield{author}{\bibinfo{person}{Xueyuan Gao} {and} \bibinfo{person}{Hua Feng}.} \bibinfo{year}{2023}\natexlab{}.
\newblock \showarticletitle{{AI}-{Driven} {Productivity} {Gains}: {Artificial} {Intelligence} and {Firm} {Productivity}}.
\newblock \bibinfo{journal}{\emph{Sustainability}} \bibinfo{volume}{15}, \bibinfo{number}{11} (\bibinfo{date}{Jan.} \bibinfo{year}{2023}), \bibinfo{pages}{8934}.
\newblock
\showISSN{2071-1050}
\urldef\tempurl%
\url{https://doi.org/10.3390/su15118934}
\showDOI{\tempurl}
\newblock
\shownote{Number: 11 Publisher: Multidisciplinary Digital Publishing Institute}.


\bibitem[Gao et~al\mbox{.}(2024)]%
        {gao_retrieval-augmented_2024}
\bibfield{author}{\bibinfo{person}{Yunfan Gao}, \bibinfo{person}{Yun Xiong}, \bibinfo{person}{Xinyu Gao}, \bibinfo{person}{Kangxiang Jia}, \bibinfo{person}{Jinliu Pan}, \bibinfo{person}{Yuxi Bi}, \bibinfo{person}{Yi Dai}, \bibinfo{person}{Jiawei Sun}, \bibinfo{person}{Meng Wang}, {and} \bibinfo{person}{Haofen Wang}.} \bibinfo{year}{2024}\natexlab{}.
\newblock \bibinfo{title}{Retrieval-{Augmented} {Generation} for {Large} {Language} {Models}: {A} {Survey}}.
\newblock
\newblock
\urldef\tempurl%
\url{https://doi.org/10.48550/arXiv.2312.10997}
\showDOI{\tempurl}
\newblock
\shownote{arXiv:2312.10997 [cs]}.


\bibitem[Gonzalez(2022)]%
        {gonzalez_council_nodate}
\bibfield{author}{\bibinfo{person}{Wendy Gonzalez}.} \bibinfo{year}{2022}\natexlab{}.
\newblock \showarticletitle{Council {Post}: {Three} {Ways} {AI} {Is} {Impacting} {The} {Automobile} {Industry}}.
\newblock \bibinfo{journal}{\emph{Forbes}} (\bibinfo{year}{2022}).
\newblock
\urldef\tempurl%
\url{https://www.forbes.com/councils/forbesbusinesscouncil/2022/04/19/three-ways-ai-is-impacting-the-automobile-industry/}
\showURL{%
\tempurl}
\newblock
\shownote{Section: Small Business}.


\bibitem[Hendrycks et~al\mbox{.}(2021)]%
        {hendrycks_measuring_2021}
\bibfield{author}{\bibinfo{person}{Dan Hendrycks}, \bibinfo{person}{Collin Burns}, \bibinfo{person}{Steven Basart}, \bibinfo{person}{Andy Zou}, \bibinfo{person}{Mantas Mazeika}, \bibinfo{person}{Dawn Song}, {and} \bibinfo{person}{Jacob Steinhardt}.} \bibinfo{year}{2021}\natexlab{}.
\newblock \bibinfo{title}{Measuring {Massive} {Multitask} {Language} {Understanding}}.
\newblock
\newblock
\urldef\tempurl%
\url{https://doi.org/10.48550/arXiv.2009.03300}
\showDOI{\tempurl}
\newblock
\shownote{arXiv:2009.03300 [cs]}.


\bibitem[Hu(2023)]%
        {hu_chatgpt_2023}
\bibfield{author}{\bibinfo{person}{Krystal Hu}.} \bibinfo{year}{2023}\natexlab{}.
\newblock \showarticletitle{{ChatGPT} sets record for fastest-growing user base - analyst note}.
\newblock \bibinfo{journal}{\emph{Reuters}} (\bibinfo{date}{Feb.} \bibinfo{year}{2023}).
\newblock
\urldef\tempurl%
\url{https://www.reuters.com/technology/chatgpt-sets-record-fastest-growing-user-base-analyst-note-2023-02-01/}
\showURL{%
\tempurl}


\bibitem[Izacard and Grave(2021)]%
        {izacard_leveraging_2021}
\bibfield{author}{\bibinfo{person}{Gautier Izacard} {and} \bibinfo{person}{Edouard Grave}.} \bibinfo{year}{2021}\natexlab{}.
\newblock \bibinfo{title}{Leveraging {Passage} {Retrieval} with {Generative} {Models} for {Open} {Domain} {Question} {Answering}}.
\newblock
\newblock
\urldef\tempurl%
\url{https://doi.org/10.48550/arXiv.2007.01282}
\showDOI{\tempurl}
\newblock
\shownote{arXiv:2007.01282 [cs]}.


\bibitem[Jeong(2024)]%
        {jeong_study_2024}
\bibfield{author}{\bibinfo{person}{Cheonsu Jeong}.} \bibinfo{year}{2024}\natexlab{}.
\newblock \bibinfo{title}{A {Study} on the {Implementation} {Method} of an {Agent}-{Based} {Advanced} {RAG} {System} {Using} {Graph}}.
\newblock
\newblock
\urldef\tempurl%
\url{https://doi.org/10.48550/arXiv.2407.19994}
\showDOI{\tempurl}
\newblock
\shownote{arXiv:2407.19994 [cs]}.


\bibitem[Kamoi et~al\mbox{.}(2024)]%
        {kamoi_evaluating_2024}
\bibfield{author}{\bibinfo{person}{Ryo Kamoi}, \bibinfo{person}{Sarkar Snigdha~Sarathi Das}, \bibinfo{person}{Renze Lou}, \bibinfo{person}{Jihyun~Janice Ahn}, \bibinfo{person}{Yilun Zhao}, \bibinfo{person}{Xiaoxin Lu}, \bibinfo{person}{Nan Zhang}, \bibinfo{person}{Yusen Zhang}, \bibinfo{person}{Ranran~Haoran Zhang}, \bibinfo{person}{Sujeeth~Reddy Vummanthala}, \bibinfo{person}{Salika Dave}, \bibinfo{person}{Shaobo Qin}, \bibinfo{person}{Arman Cohan}, \bibinfo{person}{Wenpeng Yin}, {and} \bibinfo{person}{Rui Zhang}.} \bibinfo{year}{2024}\natexlab{}.
\newblock \bibinfo{title}{Evaluating {LLMs} at {Detecting} {Errors} in {LLM} {Responses}}.
\newblock
\newblock
\urldef\tempurl%
\url{https://doi.org/10.48550/arXiv.2404.03602}
\showDOI{\tempurl}
\newblock
\shownote{arXiv:2404.03602 [cs]}.


\bibitem[Kinniment et~al\mbox{.}(2024)]%
        {kinniment_evaluating_2024}
\bibfield{author}{\bibinfo{person}{Megan Kinniment}, \bibinfo{person}{Lucas Jun~Koba Sato}, \bibinfo{person}{Haoxing Du}, \bibinfo{person}{Brian Goodrich}, \bibinfo{person}{Max Hasin}, \bibinfo{person}{Lawrence Chan}, \bibinfo{person}{Luke~Harold Miles}, \bibinfo{person}{Tao~R. Lin}, \bibinfo{person}{Hjalmar Wijk}, \bibinfo{person}{Joel Burget}, \bibinfo{person}{Aaron Ho}, \bibinfo{person}{Elizabeth Barnes}, {and} \bibinfo{person}{Paul Christiano}.} \bibinfo{year}{2024}\natexlab{}.
\newblock \bibinfo{title}{Evaluating {Language}-{Model} {Agents} on {Realistic} {Autonomous} {Tasks}}.
\newblock
\newblock
\urldef\tempurl%
\url{http://arxiv.org/abs/2312.11671}
\showURL{%
\tempurl}
\newblock
\shownote{arXiv:2312.11671 [cs]}.


\bibitem[Kwong et~al\mbox{.}(2024)]%
        {kwong_long_2024}
\bibfield{author}{\bibinfo{person}{Jethro C.~C. Kwong}, \bibinfo{person}{Serena C.~Y. Wang}, \bibinfo{person}{Grace~C. Nickel}, \bibinfo{person}{Giovanni~E. Cacciamani}, {and} \bibinfo{person}{Joseph~C. Kvedar}.} \bibinfo{year}{2024}\natexlab{}.
\newblock \showarticletitle{The long but necessary road to responsible use of large language models in healthcare research}.
\newblock \bibinfo{journal}{\emph{NPJ Digital Medicine}}  \bibinfo{volume}{7} (\bibinfo{date}{July} \bibinfo{year}{2024}), \bibinfo{pages}{177}.
\newblock
\showISSN{2398-6352}
\urldef\tempurl%
\url{https://doi.org/10.1038/s41746-024-01180-y}
\showDOI{\tempurl}


\bibitem[Lewis et~al\mbox{.}(2021)]%
        {lewis_retrieval-augmented_2021}
\bibfield{author}{\bibinfo{person}{Patrick Lewis}, \bibinfo{person}{Ethan Perez}, \bibinfo{person}{Aleksandra Piktus}, \bibinfo{person}{Fabio Petroni}, \bibinfo{person}{Vladimir Karpukhin}, \bibinfo{person}{Naman Goyal}, \bibinfo{person}{Heinrich Küttler}, \bibinfo{person}{Mike Lewis}, \bibinfo{person}{Wen-tau Yih}, \bibinfo{person}{Tim Rocktäschel}, \bibinfo{person}{Sebastian Riedel}, {and} \bibinfo{person}{Douwe Kiela}.} \bibinfo{year}{2021}\natexlab{}.
\newblock \bibinfo{title}{Retrieval-{Augmented} {Generation} for {Knowledge}-{Intensive} {NLP} {Tasks}}.
\newblock
\newblock
\urldef\tempurl%
\url{https://doi.org/10.48550/arXiv.2005.11401}
\showDOI{\tempurl}
\newblock
\shownote{arXiv:2005.11401 [cs]}.


\bibitem[Liu et~al\mbox{.}(2023)]%
        {liu_agentbench_2023}
\bibfield{author}{\bibinfo{person}{Xiao Liu}, \bibinfo{person}{Hao Yu}, \bibinfo{person}{Hanchen Zhang}, \bibinfo{person}{Yifan Xu}, \bibinfo{person}{Xuanyu Lei}, \bibinfo{person}{Hanyu Lai}, \bibinfo{person}{Yu Gu}, \bibinfo{person}{Hangliang Ding}, \bibinfo{person}{Kaiwen Men}, \bibinfo{person}{Kejuan Yang}, \bibinfo{person}{Shudan Zhang}, \bibinfo{person}{Xiang Deng}, \bibinfo{person}{Aohan Zeng}, \bibinfo{person}{Zhengxiao Du}, \bibinfo{person}{Chenhui Zhang}, \bibinfo{person}{Sheng Shen}, \bibinfo{person}{Tianjun Zhang}, \bibinfo{person}{Yu Su}, \bibinfo{person}{Huan Sun}, \bibinfo{person}{Minlie Huang}, \bibinfo{person}{Yuxiao Dong}, {and} \bibinfo{person}{Jie Tang}.} \bibinfo{year}{2023}\natexlab{}.
\newblock \bibinfo{title}{{AgentBench}: {Evaluating} {LLMs} as {Agents}}.
\newblock
\newblock
\urldef\tempurl%
\url{http://arxiv.org/abs/2308.03688}
\showURL{%
\tempurl}
\newblock
\shownote{arXiv:2308.03688 [cs]}.


\bibitem[Mandl(2023)]%
        {mandl_employment_nodate}
\bibfield{author}{\bibinfo{person}{Irene Mandl}.} \bibinfo{year}{2023}\natexlab{}.
\newblock \bibinfo{booktitle}{\emph{Employment impact of digitalisation}}.
\newblock \bibinfo{type}{{T}echnical {R}eport}. \bibinfo{institution}{Eurofund}.
\newblock
\urldef\tempurl%
\url{https://datawrapper.dwcdn.net/H37Ex/2/}
\showURL{%
\tempurl}


\bibitem[Momennejad et~al\mbox{.}(2023)]%
        {momennejad_evaluating_2023}
\bibfield{author}{\bibinfo{person}{Ida Momennejad}, \bibinfo{person}{Hosein Hasanbeig}, \bibinfo{person}{Felipe Vieira}, \bibinfo{person}{Hiteshi Sharma}, \bibinfo{person}{Robert~Osazuwa Ness}, \bibinfo{person}{Nebojsa Jojic}, \bibinfo{person}{Hamid Palangi}, {and} \bibinfo{person}{Jonathan Larson}.} \bibinfo{year}{2023}\natexlab{}.
\newblock \bibinfo{title}{Evaluating {Cognitive} {Maps} and {Planning} in {Large} {Language} {Models} with {CogEval}}.
\newblock
\newblock
\urldef\tempurl%
\url{http://arxiv.org/abs/2309.15129}
\showURL{%
\tempurl}
\newblock
\shownote{arXiv:2309.15129 [cs]}.


\bibitem[Morera and Romani(2024)]%
        {morera_metas_2024}
\bibfield{author}{\bibinfo{person}{Dani Morera} {and} \bibinfo{person}{Andre Romani}.} \bibinfo{year}{2024}\natexlab{}.
\newblock \showarticletitle{Meta's {WhatsApp} launches new {AI} tools for businesses}.
\newblock \bibinfo{journal}{\emph{Reuters}} (\bibinfo{date}{June} \bibinfo{year}{2024}).
\newblock
\urldef\tempurl%
\url{https://www.reuters.com/technology/metas-whatsapp-launches-new-ai-tools-businesses-target-messages-chats-2024-06-06/}
\showURL{%
\tempurl}


\bibitem[Nellis(2024)]%
        {nellis_mayo_2024}
\bibfield{author}{\bibinfo{person}{Stephen Nellis}.} \bibinfo{year}{2024}\natexlab{}.
\newblock \showarticletitle{Mayo {Clinic} pairs with {Cerebras} {Systems} to help develop {AI} for health care}.
\newblock \bibinfo{journal}{\emph{Reuters}} (\bibinfo{date}{Jan.} \bibinfo{year}{2024}).
\newblock
\urldef\tempurl%
\url{https://www.reuters.com/business/healthcare-pharmaceuticals/mayo-clinic-pairs-with-cerebras-systems-help-develop-ai-health-care-2024-01-09/}
\showURL{%
\tempurl}


\bibitem[Noy and Zhang(2023)]%
        {noy_experimental_2023}
\bibfield{author}{\bibinfo{person}{Shakked Noy} {and} \bibinfo{person}{Whitney Zhang}.} \bibinfo{year}{2023}\natexlab{}.
\newblock \showarticletitle{Experimental evidence on the productivity effects of generative artificial intelligence}.
\newblock \bibinfo{journal}{\emph{Science}} \bibinfo{volume}{381}, \bibinfo{number}{6654} (\bibinfo{date}{July} \bibinfo{year}{2023}), \bibinfo{pages}{187--192}.
\newblock
\urldef\tempurl%
\url{https://doi.org/10.1126/science.adh2586}
\showDOI{\tempurl}
\newblock
\shownote{Publisher: American Association for the Advancement of Science}.


\bibitem[O'Brien(2024)]%
        {obrien_future_2024}
\bibfield{author}{\bibinfo{person}{Keith O'Brien}.} \bibinfo{year}{2024}\natexlab{}.
\newblock \bibinfo{title}{The {Future} of {AI} in {Customer} {Service}}.
\newblock
\newblock
\urldef\tempurl%
\url{https://www.ibm.com/think/insights/customer-service-future}
\showURL{%
\tempurl}


\bibitem[Perrone(2024)]%
        {perrone_will_2024}
\bibfield{author}{\bibinfo{person}{Matthew Perrone}.} \bibinfo{year}{2024}\natexlab{}.
\newblock \showarticletitle{Will {AI} replace doctors who read {X}-rays, or just make them better than ever?}
\newblock \bibinfo{journal}{\emph{AP News}} (\bibinfo{date}{May} \bibinfo{year}{2024}).
\newblock
\urldef\tempurl%
\url{https://apnews.com/article/ai-algorithms-chatgpt-doctors-radiologists-3bc95db51a41469c390b0f1f48c7dd4e}
\showURL{%
\tempurl}
\newblock
\shownote{Section: Health}.


\bibitem[Pogla(2024)]%
        {pogla_auto-gpt_2024}
\bibfield{author}{\bibinfo{person}{Matt Pogla}.} \bibinfo{year}{2024}\natexlab{}.
\newblock \bibinfo{title}{Auto-{GPT} vs {ChatGPT}: {How} {Do} {They} {Differ}}.
\newblock
\newblock
\urldef\tempurl%
\url{https://autogpt.net/auto-gpt-vs-chatgpt-how-do-they-differ-and-everything-you-need-to-know/}
\showURL{%
\tempurl}


\bibitem[Rein et~al\mbox{.}(2023)]%
        {rein_gpqa_2023}
\bibfield{author}{\bibinfo{person}{David Rein}, \bibinfo{person}{Betty~Li Hou}, \bibinfo{person}{Asa~Cooper Stickland}, \bibinfo{person}{Jackson Petty}, \bibinfo{person}{Richard~Yuanzhe Pang}, \bibinfo{person}{Julien Dirani}, \bibinfo{person}{Julian Michael}, {and} \bibinfo{person}{Samuel~R. Bowman}.} \bibinfo{year}{2023}\natexlab{}.
\newblock \bibinfo{title}{{GPQA}: {A} {Graduate}-{Level} {Google}-{Proof} {Q}\&{A} {Benchmark}}.
\newblock
\newblock
\urldef\tempurl%
\url{https://doi.org/10.48550/arXiv.2311.12022}
\showDOI{\tempurl}
\newblock
\shownote{arXiv:2311.12022 [cs]}.


\bibitem[Saha(2022)]%
        {saha_help_nodate}
\bibfield{author}{\bibinfo{person}{Sudip Saha}.} \bibinfo{year}{2022}\natexlab{}.
\newblock \bibinfo{title}{Help {Desk} {Software} {Market} {Size}, {Shares} \& {Forecast} – 2032}.
\newblock
\newblock
\urldef\tempurl%
\url{https://www.futuremarketinsights.com/reports/help-desk-software-market}
\showURL{%
\tempurl}


\bibitem[Shridhar et~al\mbox{.}(2020)]%
        {shridhar_alfred_2020}
\bibfield{author}{\bibinfo{person}{Mohit Shridhar}, \bibinfo{person}{Jesse Thomason}, \bibinfo{person}{Daniel Gordon}, \bibinfo{person}{Yonatan Bisk}, \bibinfo{person}{Winson Han}, \bibinfo{person}{Roozbeh Mottaghi}, \bibinfo{person}{Luke Zettlemoyer}, {and} \bibinfo{person}{Dieter Fox}.} \bibinfo{year}{2020}\natexlab{}.
\newblock \bibinfo{title}{{ALFRED}: {A} {Benchmark} for {Interpreting} {Grounded} {Instructions} for {Everyday} {Tasks}}.
\newblock
\newblock
\urldef\tempurl%
\url{http://arxiv.org/abs/1912.01734}
\showURL{%
\tempurl}
\newblock
\shownote{arXiv:1912.01734 [cs]}.


\bibitem[{Stack Exchange, Inc.}({[n.\,d.]})]%
        {stack_exchange_inc_stack_nodate}
\bibfield{author}{\bibinfo{person}{{Stack Exchange, Inc.}}} \bibinfo{year}{[n.\,d.]}\natexlab{}.
\newblock \bibinfo{title}{Stack {Exchange} {Data} {Dump}}.
\newblock
\newblock
\urldef\tempurl%
\url{https://archive.org/download/stackexchange/superuser.com.7z}
\showURL{%
\tempurl}


\bibitem[Tsymbal(2024)]%
        {tsymbal_ai_2024}
\bibfield{author}{\bibinfo{person}{Tetiana Tsymbal}.} \bibinfo{year}{2024}\natexlab{}.
\newblock \bibinfo{title}{{AI} in {Customer} {Service} {Statistics} [{April} 2024]}.
\newblock
\newblock
\urldef\tempurl%
\url{https://masterofcode.com/blog/ai-in-customer-service-statistics}
\showURL{%
\tempurl}


\bibitem[Williams and Huckle(2024)]%
        {williams_easy_2024}
\bibfield{author}{\bibinfo{person}{Sean Williams} {and} \bibinfo{person}{James Huckle}.} \bibinfo{year}{2024}\natexlab{}.
\newblock \bibinfo{title}{Easy {Problems} {That} {LLMs} {Get} {Wrong}}.
\newblock
\newblock
\urldef\tempurl%
\url{https://doi.org/10.48550/arXiv.2405.19616}
\showDOI{\tempurl}
\newblock
\shownote{arXiv:2405.19616 [cs]}.


\bibitem[Wu et~al\mbox{.}(2024)]%
        {wu_smartplay_2024}
\bibfield{author}{\bibinfo{person}{Yue Wu}, \bibinfo{person}{Xuan Tang}, \bibinfo{person}{Tom~M. Mitchell}, {and} \bibinfo{person}{Yuanzhi Li}.} \bibinfo{year}{2024}\natexlab{}.
\newblock \bibinfo{title}{{SmartPlay}: {A} {Benchmark} for {LLMs} as {Intelligent} {Agents}}.
\newblock
\newblock
\urldef\tempurl%
\url{https://doi.org/10.48550/arXiv.2310.01557}
\showDOI{\tempurl}
\newblock
\shownote{arXiv:2310.01557 [cs]}.


\bibitem[Xu et~al\mbox{.}(2024)]%
        {xu_large_2024}
\bibfield{author}{\bibinfo{person}{Derong Xu}, \bibinfo{person}{Wei Chen}, \bibinfo{person}{Wenjun Peng}, \bibinfo{person}{Chao Zhang}, \bibinfo{person}{Tong Xu}, \bibinfo{person}{Xiangyu Zhao}, \bibinfo{person}{Xian Wu}, \bibinfo{person}{Yefeng Zheng}, \bibinfo{person}{Yang Wang}, {and} \bibinfo{person}{Enhong Chen}.} \bibinfo{year}{2024}\natexlab{}.
\newblock \bibinfo{title}{Large {Language} {Models} for {Generative} {Information} {Extraction}: {A} {Survey}}.
\newblock
\newblock
\urldef\tempurl%
\url{https://doi.org/10.48550/arXiv.2312.17617}
\showDOI{\tempurl}
\newblock
\shownote{arXiv:2312.17617 [cs]}.


\bibitem[Zara(2024)]%
        {zara_ai_2024}
\bibfield{author}{\bibinfo{person}{Christopher Zara}.} \bibinfo{year}{2024}\natexlab{}.
\newblock \bibinfo{title}{{AI} chatbots are taking over customer service, but most of us would rather wait for a human}.
\newblock
\newblock
\urldef\tempurl%
\url{https://www.fastcompany.com/91115626/ai-chatbots-vs-humans-customer-service-tasks-survey-ratings}
\showURL{%
\tempurl}


\bibitem[Zhang et~al\mbox{.}(2024)]%
        {zhang_raft_2024}
\bibfield{author}{\bibinfo{person}{Tianjun Zhang}, \bibinfo{person}{Shishir~G. Patil}, \bibinfo{person}{Naman Jain}, \bibinfo{person}{Sheng Shen}, \bibinfo{person}{Matei Zaharia}, \bibinfo{person}{Ion Stoica}, {and} \bibinfo{person}{Joseph~E. Gonzalez}.} \bibinfo{year}{2024}\natexlab{}.
\newblock \bibinfo{title}{{RAFT}: {Adapting} {Language} {Model} to {Domain} {Specific} {RAG}}.
\newblock
\newblock
\urldef\tempurl%
\url{https://doi.org/10.48550/arXiv.2403.10131}
\showDOI{\tempurl}
\newblock
\shownote{arXiv:2403.10131 [cs]}.


\end{thebibliography}

\clearpage
\appendix
\section*{Appendix}
\renewcommand{\thesection}{\Alph{section}} 
\section{Comparison of Groups}\label{app}

\begin{table}[ht]
    \centering
    \caption*{Comparison of Groups}
    \begin{tabular}{@{}l l r r r r r@{}}
        \toprule
        Group 1 & Group 2 & Mean Diff & p-adj & Lower & Upper & Reject \\
        \midrule
        gpt-4-1106-preview+RAG & gpt-4-1106-preview & -5.9471 & 0 & -7.2174 & -4.6767 & TRUE \\
        gpt-4-1106-preview+RAG & human & -6.2272 & 0 & -7.5017 & -4.9528 & TRUE \\
        gpt-4-1106-preview+RAG & mistral-7b+RAG & -0.8881 & 0.3833 & -2.1409 & 0.3646 & FALSE \\
        gpt-4-1106-preview+RAG & mistral-7b & -1.2711 & 0.046 & -2.5304 & -0.0119 & TRUE \\
        gpt-4-1106-preview+RAG & mixtral-8x7b+RAG & -0.5611 & 0.8557 & -1.7723 & 0.6502 & FALSE \\
        gpt-4-1106-preview+RAG & mixtral-8x7b & -5.7015 & 0 & -6.9419 & -4.4611 & TRUE \\
        gpt-4-1106-preview+RAG & mixtral-8x7b & -1.6537 & 0.0013 & -2.8918 & -0.4156 & TRUE \\
        gpt-4-1106-preview & human & -0.2802 & 0.9979 & -1.5605 & 1.0002 & FALSE \\
        gpt-4-1106-preview & mistral-7b+RAG & 5.0589 & 0 & 3.8002 & 6.3177 & TRUE \\
        gpt-4-1106-preview & mistral-7b & 4.6759 & 0 & 3.4107 & 5.9411 & TRUE \\
        gpt-4-1106-preview & mixtral-8x7b+RAG & 5.386 & 0 & 4.1686 & 6.6034 & TRUE \\
        gpt-4-1106-preview & mixtral-8x7b & 0.2455 & 0.9989 & -1.0009 & 1.4919 & FALSE \\
        gpt-4-1106-preview & mixtral-8x7b & 4.2934 & 0 & 3.0492 & 5.5375 & TRUE \\
        human & mistral-7b+RAG & 5.3391 & 0 & 4.0763 & 6.602 & TRUE \\
        human & mistral-7b & 4.9561 & 0 & 3.6867 & 6.2254 & TRUE \\
        human & mixtral-8x7b+RAG & 5.6662 & 0 & 4.4445 & 6.8879 & TRUE \\
        human & mixtral-8x7b & 0.5257 & 0.9085 & -0.7249 & 1.7763 & FALSE \\
        human & mixtral-8x7b & 4.5735 & 0 & 3.3252 & 5.8219 & TRUE \\
        mistral-7b+RAG & mistral-7b & -0.383 & 0.9831 & -1.6306 & 0.8645 & FALSE \\
        mistral-7b+RAG & mixtral-8x7b+RAG & 0.3271 & 0.9916 & -0.872 & 1.5261 & FALSE \\
        mistral-7b+RAG & mixtral-8x7b & -4.8134 & 0 & -6.0419 & -3.5849 & TRUE \\
        mistral-7b+RAG & mixtral-8x7b & -0.7656 & 0.5559 & -1.9918 & 0.4606 & FALSE \\
        mistral-7b & mixtral-8x7b+RAG & 0.7101 & 0.6303 & -0.4958 & 1.9159 & FALSE \\
        mistral-7b & mixtral-8x7b & -4.4304 & 0 & -5.6655 & -3.1953 & TRUE \\
        mistral-7b & mixtral-8x7b & -0.3826 & 0.9821 & -1.6154 & 0.8503 & FALSE \\
        mixtral-8x7b+RAG & mixtral-8x7b & -5.1405 & 0 & -6.3266 & -3.9544 & TRUE \\
        mixtral-8x7b+RAG & mixtral-8x7b & -1.0927 & 0.0954 & -2.2764 & 0.0911 & FALSE \\
        mixtral-8x7b & mixtral-8x7b & 4.0478 & 0 & 2.8343 & 5.2613 & TRUE \\
        \bottomrule
    \end{tabular}
    \label{tab:comparison}
\end{table}

\end{document}